\documentclass[conf]{IEEEtran}
\usepackage[utf8]{inputenc}
\usepackage{amsmath}
\usepackage{cite}
\usepackage{multicol}
\usepackage{graphicx}
\usepackage{array}
\usepackage{multirow}
\usepackage{graphicx}
\usepackage{enumitem}
\usepackage{color}
\usepackage{nomencl}
\usepackage{algorithm}
\usepackage{algpseudocode}
\usepackage{ifthen}
\usepackage{etoolbox}
\usepackage{amssymb}
\usepackage{comment}
\usepackage{mathrsfs}
\usepackage{mathtools}
\usepackage{enumitem}
\usepackage{bbold}
\usepackage{yfonts}
\usepackage{soul}
\usepackage{placeins}
\usepackage{fnpct}
\usepackage{courier}
\usepackage{subfiles}
\usepackage{ntheorem}
\theoremseparator{:}

\usepackage{url}
\usepackage{hyperref}

\usepackage{fnpct}
\ifCLASSOPTIONcompsoc \usepackage[caption=false,font=normalsize,labelfon
t=sf,textfont=sf]{subfig}
\else
\usepackage[caption=false,font=footnotesize]{subfig}
\fi
\usepackage{tikz}
\usetikzlibrary{decorations.pathreplacing}
\usetikzlibrary{arrows,shapes,positioning}
\usetikzlibrary{patterns}
\usepackage{pgfplots}
\pgfplotsset{compat=newest}
\pgfplotsset{plot coordinates/math parser=false}
\newlength\figureheight
\newlength\matlabfigurewidth
\title{Experimental Validation of Model-less Robust Voltage Control using Measurement-based Estimated Voltage Sensitivity Coefficients}

\author{\IEEEauthorblockN{Rahul Gupta, Mario Paolone}\\
\IEEEauthorblockA{\textit{Distributed Electrical Systems Laboratory}},
\textit{Ecole Polytechnique Federale de Lausanne, Switzerland}\\
\{rahul.gupta, mario.paolone\}@epfl.ch}
   
\usepackage{etoolbox}
\makeatletter
\patchcmd{\@maketitle}
  {\addvspace{0.5\baselineskip}\egroup}
  {\addvspace{-2\baselineskip}\egroup}
  {}
  {}
\makeatother
\begin{document}
\markboth{PowerTech Belgrade 2023,  June 25-29 (Accepted for presentation).}%
{}
\maketitle
\begin{abstract}
Increasing adoption of smart meters and phasor measurement units (PMUs) in power distribution networks are enabling the adoption of data-driven/model-less control schemes to mitigate grid issues such as over/under voltages and power-flow congestions. However, such a scheme can lead to infeasible/inaccurate control decisions due to measurement inaccuracies. In this context, the authors' previous work proposed a robust measurement-based control scheme accounting for the uncertainties of the estimated models. In this scheme, a recursive least squares (RLS)-based method estimates the grid model (in the form of voltage magnitude sensitivity coefficients).
Then, a robust control problem optimizes power set-points of distributed energy resources (DERs) such that the nodal voltage limits are satisfied. The estimated voltage sensitivity coefficients are used to model the nodal voltages, and the control robustness is achieved by accounting for their uncertainties. 
This work presents the first experimental validation of such a robust model-less control scheme on a real power distribution grid. The scheme is applied for voltage control by regulating two photovoltaic (PV) inverters connected in a real microgrid which is a replica of the CIGRE benchmark microgrid network at the EPFL Distributed Electrical Systems Laboratory. 
\end{abstract}
\vspace{-0.8em}
\section{Introduction}
\textcolor{black}{Distribution System Operators (DSOs) have the responsibility of maintaining adequate voltage quality for end-consumers through the operation of their networks \cite{guide2004voltage}. To this end, voltage control has been identified as a prominent approach to be employed \cite{hatziargyrioucigre,cigre2011c6, pilo2012planning}. Various control strategies have been proposed in the literature, which can be broadly classified into two categories. The first category comprises the \emph{model-based control} schemes, such as those proposed in \cite{agalgaonkar2013distribution, gupta2020grid, christakou2013efficient}, which rely on accurate knowledge of the grid model, including its topology and branch parameters.}
However, it is not applicable to cases when the grid model is unavailable. To tackle this issue, \emph{data-driven} methods have been recently proposed \cite{mugnier2016model, su2019augmented, carpita2019low, valverde2018estimation, da2019data, nowak2020measurement, gupta2021compound, gupta2022model} where the network model is first inferred from suitable measurements, then fed to a coupled control scheme. These schemes are referred to as \emph{model-less control}. 
The works in \cite{mugnier2016model, carpita2019low} proposed voltage control using estimated voltage sensitivity coefficients. However, as shown in \cite{valverde2018estimation}, these schemes might suffer from the multi-collinearity problem (i.e., unreliable estimates in the case of similar power injections at multiple nodes). To solve this issue, it proposed adopting the principle component analysis (PCA) method. The works in \cite{da2019data, gupta2022model} used a two-stage estimation scheme where a least-squares (LS) method obtains a rough estimate of the sensitivity coefficients that are then corrected via an online recursive-least-square (RLS)-based scheme using the most recent measurements.

In all the above schemes \cite{mugnier2016model, carpita2019low, valverde2018estimation, da2019data}, the sensitivity coefficients were modeled as point estimates, ignoring the estimation uncertainty, which may lead to infeasible control decisions. Since measurement-based estimates are sensitive to measurement quality (i.e., bias and noise), it might be helpful to account for these elements via uncertainty bounds on the estimates in the control problem.
Author's previous work \cite{gupta2022model, GuptaThesis} proposed a robust voltage control scheme that modeled sensitivity coefficient estimates as probability density functions (PDFs) instead of their mean value. This approach resulted in a better performance than its non-robust counterpart (i.e., when the estimates were modeled by their mean values).
However, to the author's best knowledge, none of these schemes were experimentally validated. 

In this context, this paper presents an experimental validation of the estimation and control scheme of \cite{gupta2022model} on a real-scale microgrid hosted at the EPFL Distributed Electrical Systems Laboratory. The microgrid is a low-voltage distribution network and is a replica of the CIGRE microgrid benchmark network \cite{papathanassiou2005benchmark}; it hosts two\footnote{Note that the power injected from multiple PV plants may introduce the problem of multi-collinearity in the sensitivity estimation problem.} controllable photovoltaic (PV) inverters and multiple uncontrollable injections.
The grid is equipped with seven phasor measurement units (PMUs) and a dedicated communication network providing real-time measurements of nodal voltages and branch current flows. 

The paper is organized as follows: Section~\ref{sec:prob_stat} presents the problem statement, Section~\ref{sec:methods} describes the estimation and control problem, Section~\ref{sec:expt_val} presents the experimental results, and Section~\ref{sec:conclusion} summarizes the main contributions of this paper.
\vspace{-2em}
\begin{figure*}[!ht]
    \centering
    \includegraphics[width = 0.83\linewidth]{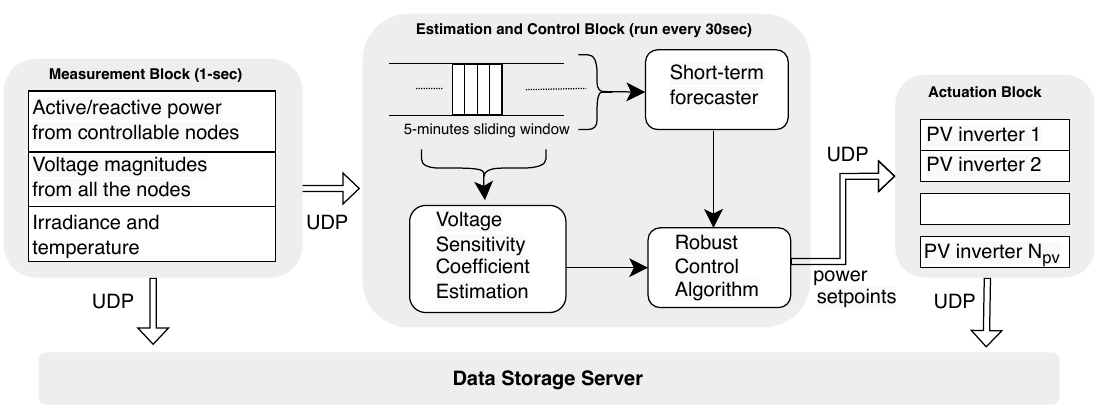}
    \caption{Schematic diagram of the model-less/measurement-based robust voltage control framework.}
    \label{fig:flow}
\end{figure*}
\section{Problem Statement}\label{sec:prob_stat}
We consider a power distribution network of generic topology (i.e., meshed or radial) equipped with measurement devices (either smart meters or phasor measurement units) capable of providing high throughput measurements\footnote{\textcolor{black}{Future distribution grids are recommended to install low-cost PMUs for their situational awareness by CIGRE and IEEE working groups \cite{d2009global, pilo2012planning}, providing high throughput measurements.}} on nodal voltage magnitudes and active/reactive powers. The objective is to control distributed energy resources (DERs) in a power distribution grid such that nodal voltages are kept within their operational bounds. We rely on the measurement-based estimation and robust voltage control originally proposed in \cite{gupta2022model}. 
The scheme is schematically shown in Fig.~\ref{fig:flow}. It consists of three different blocks: 
\begin{itemize}[leftmargin=*]
    \item \textbf{Measurement:} \textcolor{black}{the measurements of nodal active, reactive powers, and voltage magnitudes are obtained from PMUs. Also, the measurements on global horizontal irradiance (GHI) and temperature} is gathered. \textcolor{black}{We consider the measurements sampled at 1 second.}
    \item \textbf{Estimation and control:} the voltage magnitude sensitivity coefficients are estimated using a pre-defined measurement window (5-minutes in this case), \textcolor{black}{the short-term forecasts of the injections are updated, and finally, using the estimated sensitivity coefficients and short-term forecasts, the robust control algorithm computes the active and reactive power setpoints of controllable DERs such that the system state is feasible with respect to the sensitivity coefficients uncertainties.} This block is run every 30 seconds.
    \item \textbf{Actuation:} \textcolor{black}{the power setpoints are sent to the DERs through a dedicated IPv4 communication network via User Datagram Protocol (UDP).}
\end{itemize}
This scheme is experimentally validated on a real microgrid which is a replica of the CIGRE benchmark microgrid network hosted at the EPFL Distributed Electrical Systems Laboratory. The framework is validated in terms of estimation and control accuracy. In the following, we briefly recall the control and estimation scheme.
\section{Methods}\label{sec:methods}
Let us consider a power distribution network consisting of $N_b$ non-slack buses contained in the set $\mathcal{N}_b = \{1,\dots, N_b\}$. The distribution network hosts multiple DERs that can be controlled {to provide} active and reactive power support to the grid. The objective is to control a subset of these DERs at a relatively high refresh rate (e.g., 30 seconds) such that nodal voltage constraints are always satisfied. In the following, we present how the nodal voltages are modeled using measurements, the model-less robust control problem, and the estimation problem formulations.

The nodal voltages are approximated via the first-order Taylor's approximation using the so-called voltage sensitivity coefficients. Let $|{v}_{i,t_{k-1}}| \in \mathbb{R}$ denote the nodal voltage magnitude of $i-$th node at time $t_{k-1}$, ${\mathbf{K}}^p_{i,t_{k-1}}\in \mathbb{R}^{N_b\times 1}$ and ${\mathbf{K}}^q_{i,t_{k-1}}\in \mathbb{R}^{N_b\times 1}$ be the nodal voltage magnitude sensitivity coefficient with respect to nodal injections of active and reactive powers, respectively. 
Let the nodal active and reactive power injections are denoted by $\mathbf{p}_{t_k} \in \mathbb{R}^{1\times N_b}$ and $\mathbf{q}_{t_k} \in \mathbb{R}^{1\times N_b}$, respectively; $\mathbf{p}_{t_k} - \mathbf{p}_{t_{k-1}} = \Delta\mathbf{p}_{t_k}, \mathbf{q}_{t_k} - \mathbf{q}_{t_{k-1}} = \Delta\mathbf{q}_{t_k}$ be the variations of nodal active and reactive power injections. The nodal voltage magnitude of $i-$th node at time $t_{k}$ (i.e., $|{v}_{i,t_{k}}|$) can be approximated by
\begin{align}
\small \hspace{-0.8em}
    |{v}_{i,t_{k}}| \approx |{v}_{i,t_{k-1}}| +  & \Delta{\mathbf{p}_{t_k}}{\mathbf{K}}^p_{i,t_{k-1}}+ 
    \Delta{\mathbf{q}_{t_k}}{\mathbf{K}}^q_{i,t_{k-1}}  ~ \forall i \in \mathcal{N}_b \label{eq:volt_model}
\end{align}
To account for the uncertainty on the estimates, the coefficients are represented by following intervals with $\Delta \mathbf{K}^p_{i,t_k}, \Delta \mathbf{K}^q_{i,t_k}$ being the estimated uncertainty
\begin{subequations}
\label{eq:intervalCoeff}
\begin{align}
    & \mathbf{K}^p_{i,{t_k}} \in [\widehat{\mathbf{K}}^p_{i,{t_k}}- \Delta \mathbf{K}^p_{i,{t_k}}, ~\widehat{\mathbf{K}}^p_{i,{t_k}}+ \Delta \mathbf{K}^p_{i,{t_k}}] \label{eq:v1} && \forall i \in \mathcal{N}_b\\
    & \mathbf{K}^q_{i,{t_k}} \in [\widehat{\mathbf{K}}^q_{i,{t_k}}- \Delta \mathbf{K}^q_{i,{t_k}}, ~\widehat{\mathbf{K}}^q_{i,{t_k}}+ \Delta \mathbf{K}^q_{i,{t_k}}] && \forall i \in \mathcal{N}_b \label{eq:v2}.
\end{align}
\end{subequations}

\textcolor{black}{The sensitivity coefficients in \eqref{eq:volt_model} are obtained by a measurement-based estimation process described later in Appendix~\ref{sec:sensitivity_Est}.
We use a two-stage scheme from \cite{gupta2022model, GuptaThesis} for the estimation of the sensitivity coefficients that are used as inputs in the robust control problem of the previous section. 
In the first stage, the least squares (LS)-based scheme (offline) estimates the initial values of sensitivity coefficients using previous-day measurements. Then, an online recursive least squares (RLS) scheme is used to update the estimates during real-time operation.  \textcolor{black}{The scheme is described in Appendix~\ref{sec:sensitivity_Est}.}}
\subsection{Robust Control Problem Formulation}
\label{sec:Robust_control}
\textcolor{black}{We refer to the robust control formulation from the author's previous work in \cite{gupta2022model, GuptaThesis}. Here, the objective is to control active/reactive power injections from curtailable PV plants such that the grid nodal voltages are always within the operational bounds. At the same time, it minimizes the active power curtailment with respect to maximum power potential (MPP) and the corresponding reactive power constrained by the PV plant's minimum power factor. The control formulation is linear, thanks to the linear grid constraints modeled by voltage sensitivity coefficients and robust reformulation scheme \cite{bertsimas2011theory} and is briefly described in Appendix~\ref{sec:Robust_control_append}.}
\begin{figure*}[!ht]
    \centering
    \subfloat[]{\includegraphics[width=0.43\linewidth]{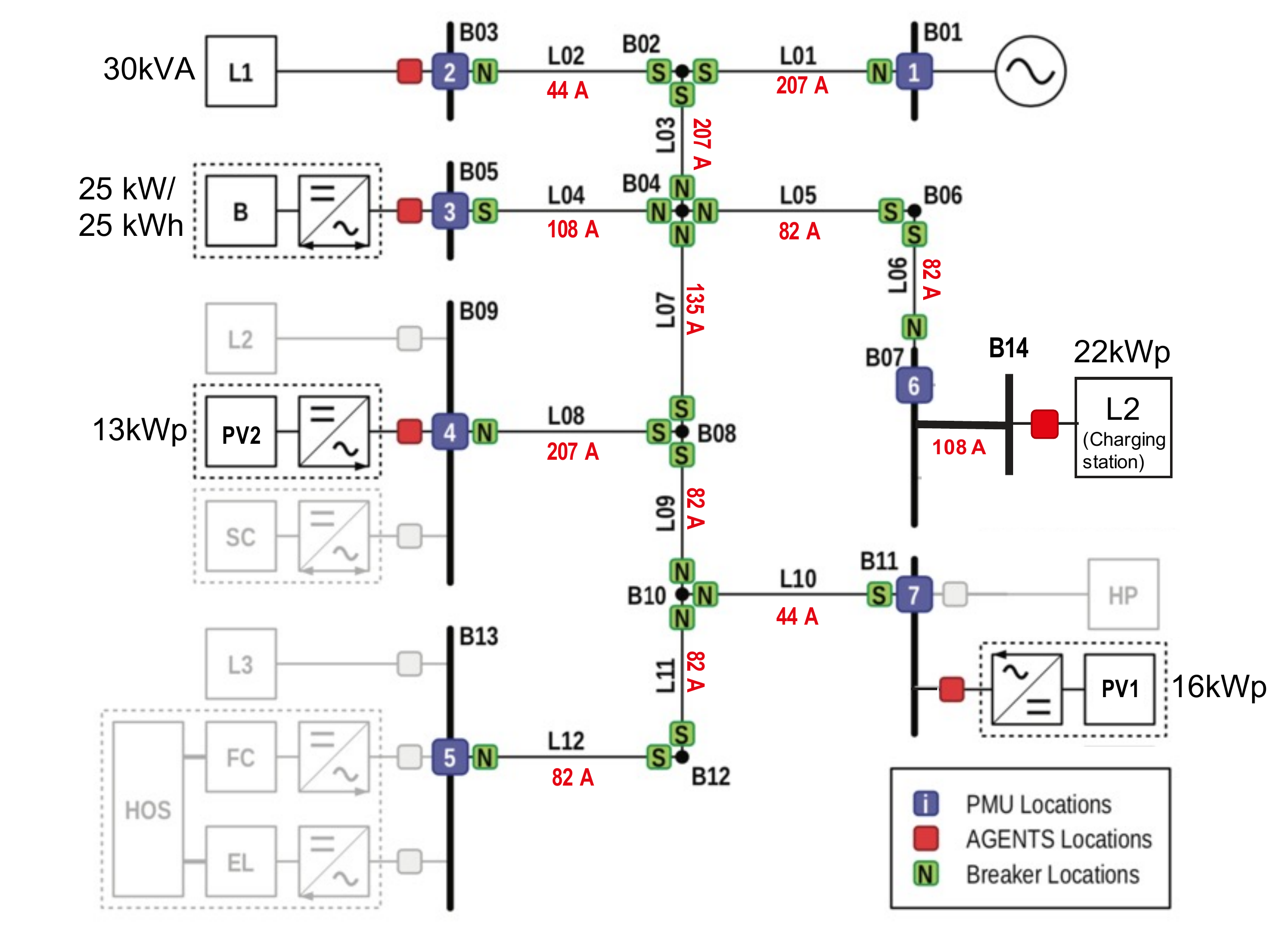}\label{fig:microgrid}}
    \subfloat[]{\includegraphics[width=0.21\linewidth]{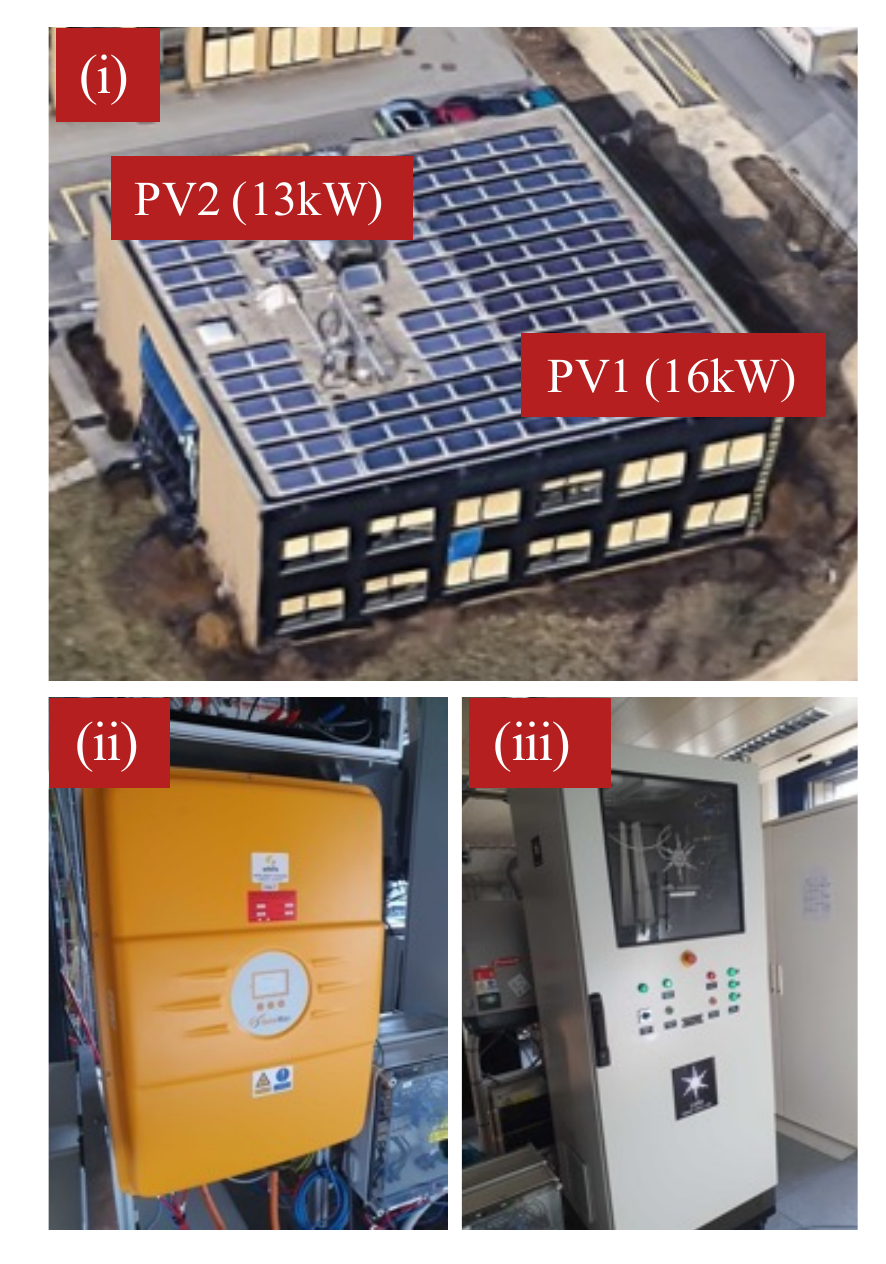}\label{fig:pvplant}}
    \subfloat[]{\includegraphics[width=0.36\linewidth]{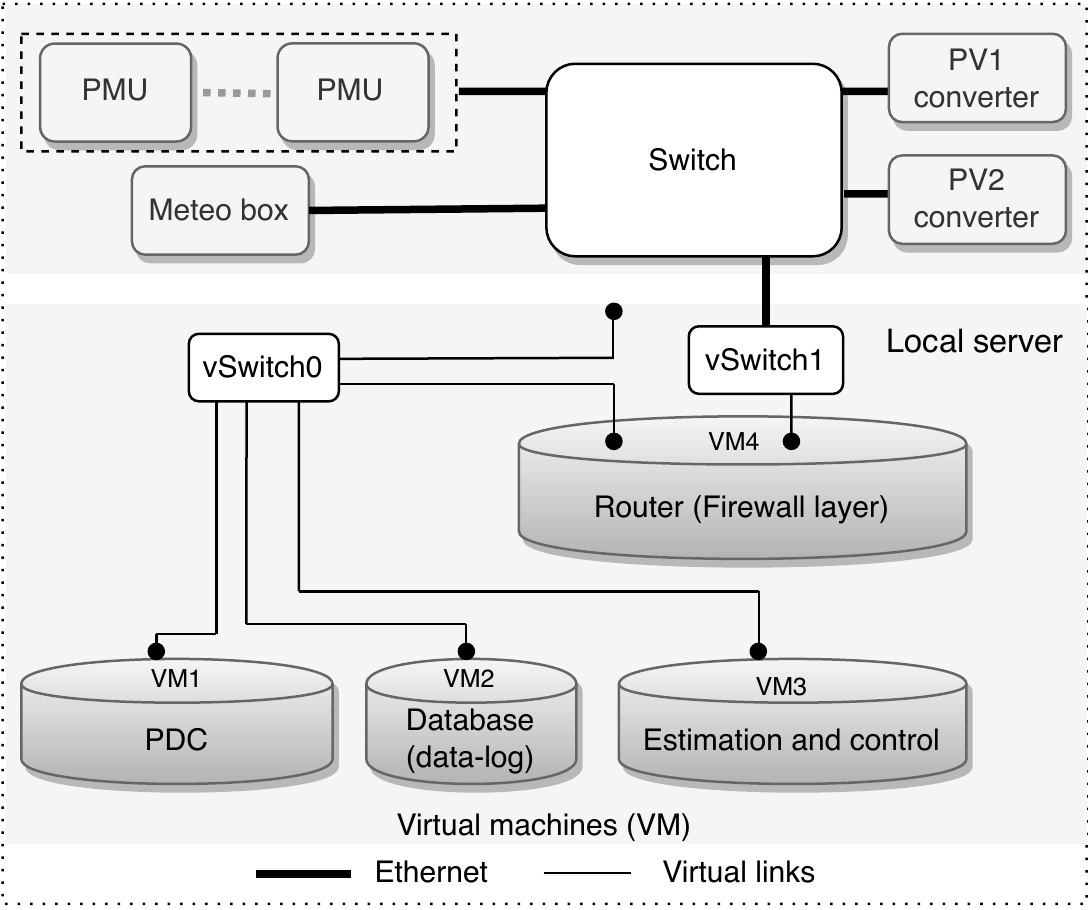}\label{fig:DESLserver}}
    \label{fig:my_label}
    \caption{(a) The microgrid setup used for the experimental validation. We consider two controllable resources (curtailable PV plants PV1 and PV2 at buses B11 and B09) and uncontrollable injections from L1, B, and L2 at buses B03, B05, and B14, respectively. (b-i) The two PV plants at the roof-top, (b-ii, iii) inverters for PV1 and PV2, respectively, and (c) IT communication infrastructure for microgrid including local server, PMUs, Meteo-box and PV converters.}
\end{figure*}
\section{Experimental Validation}\label{sec:expt_val}
In the following, we present the experimental validation of the measurement-based estimation and robust control scheme. First, we describe the experimental setup, and then the results from a full day of experiments are presented and discussed.
\subsection{Experimental Setup}
\begin{figure}[!ht]
    \centering
    \includegraphics[width=0.7\linewidth]{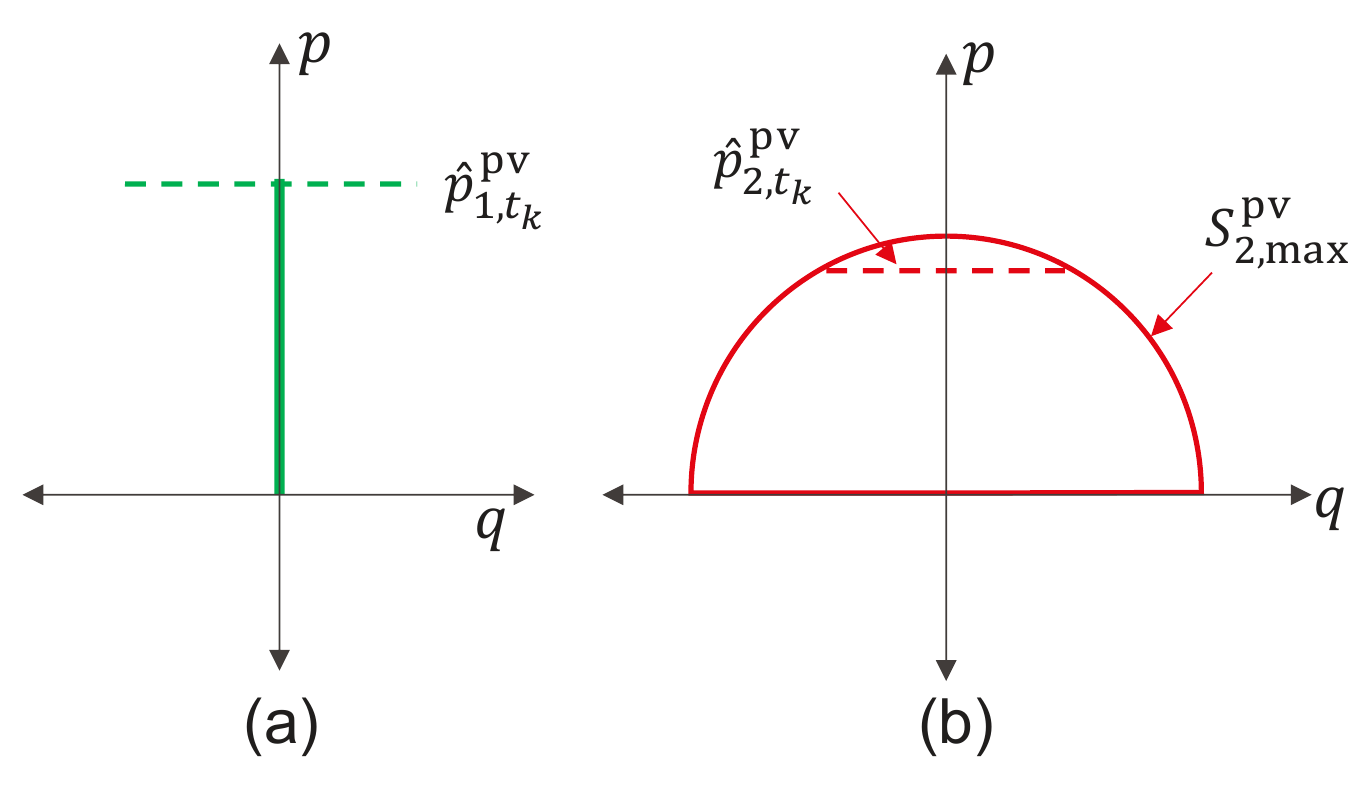}
    \caption{Capability curves of (a) PV1 and (b) PV2 converters.}
    \label{fig:cap_curve}
\end{figure}

We validate the model-less robust control scheme on a real-scale microgrid hosted at the EPFL's Distributed Electrical Systems Laboratory. The microgrid setup is a replica of the CIGRE low voltage benchmark microgrid \cite{papathanassiou2005benchmark}. The grid topology, corresponding ampacities, and locations of the DERs are shown in Fig.~\ref{fig:microgrid}. It also shows the locations of the PMUs installed at nodes B01, B03, B05, B09, B13, B07, and B11, respectively. It is operated at 400~V and is connected to the 20~kV medium voltage feeder via a 630~kVA transformer. The microgrid hosts several controllable and uncontrollable DERs. For the sake of this experiment, we consider two PV plants as controllable, other injections from Battery (B), loads (L1 and L2) are used to imitate uncontrollable prosumers. The nominal ratings of DERs are also displayed in Fig.~\ref{fig:microgrid}. The PV panels are shown in Fig.~(\ref{fig:pvplant}-i), and they are interfaced by two different converters (shown in Fig.~\ref{fig:pvplant}-ii and iii) at nodes B09 and B11, respectively. The two converters differ in terms of their controllability. Their capability curves are shown in Fig.~\ref{fig:cap_curve}. From the curve, it is apparent that the first converter cannot control reactive power, whereas the second can control both the active and reactive powers within its capacity. In both cases, the active power is limited by the MPP obtained using a short-term irradiance forecast.
\subsubsection{Monitoring and Communication Infrastructure}\label{sec:monitoring}
The microgrid is equipped with seven PMUs providing measurements of nodal voltages and lines currents. Although the measurements are available at a time sampling of 100~ms, we down-sample them to 1~second to demonstrate that the proposed scheme can work with that time resolution. Since the estimation approach requires power measurements, we compute the nodal powers using the voltage and current measurements from PMUs. The nodal voltages and branch currents are sensed by commercial LEM voltage transducer CV 3-1000 \cite{V_IT_DESL} and current transducer LF 205-S/SP3 \cite{I_IT_DESL} with IT measurement classes of 0.2 and 0.5, respectively. We also obtain global horizontal irradiance (GHI) and air temperature measurements from suitable meteo-boxex that allows obtaining a short-term forecast of the PV generation. The specifications of the PMUs and meteo-boxex are described in \cite{gupta2022reliable}.

As for communication, the microgrid is equipped with a dedicated IPv4 communication infrastructure. The network layout is shown in Fig.~\ref{fig:DESLserver}. It connects the PMUs, meteo-boxes, and a local server. The local server hosts four different virtual machines (VM), their functions are (i) VM1 as PDC (phasor data concentrator) for aggregating the packets from the PMUs, (ii) VM2: Databases for logging the measurements, (ii) VM3: Implementing the estimation and control algorithms and (iv) VM4: Routing the packets according to the firewall configurations. We use User Datagram Protocol (UDP) for the communication of the packets. All the elements are connected to the sub-network via Ethernet cables.  

\subsubsection{Short-term forecast}
Thanks to the fast time resolution (1-second) of the measurements, we use persistent\footnote{A better forecasting strategy will be investigated in future work.} forecasting scheme for short-term forecasts of GHI and demand. In other words, we assume that the GHI and demand stay the same (as observed in the last second) for the computation of the control setpoint. 
The GHI and air temperature measurements are obtained via meteo-box described in Sec.~\ref{sec:monitoring}. These measurements are then used for the estimation of the PV MPP generation using a PV model from \cite{sossan2019solar}. 
\subsubsection{Experimental flow diagram}
\begin{figure}
        \includegraphics[width=\linewidth]{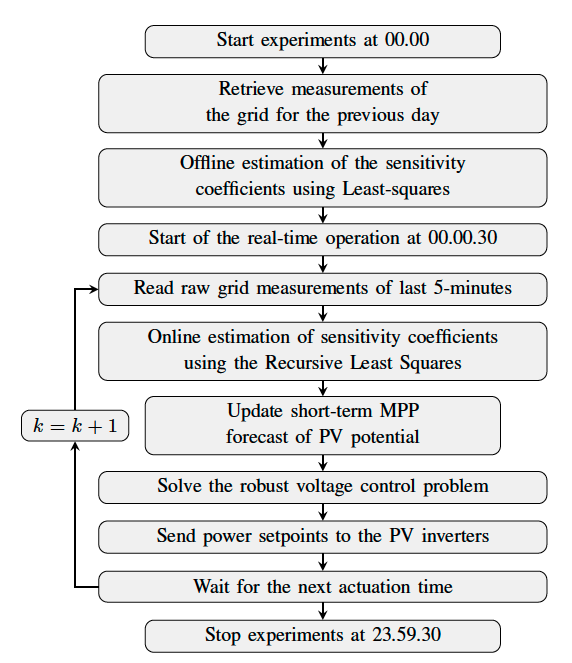}
        \caption{Flow-chart illustrating the real-time operation for a single day of operation.}
    \label{fig:modelless_RTflow}
\end{figure}

Fig.~\ref{fig:modelless_RTflow} shows the data flow during the real-time operation. It starts at midnight 00.00 UTC. As described in Appendix~\ref{sec:sensitivity_Est}, the initial sensitivity coefficients are estimated by LS using the previous day's measurements. Then, the real-time stage starts at 00.00.30. Using the recent 5-minute measurements (sampled per second), it updates the estimation of the sensitivity coefficients using RLS. In the next step, we update the MPP and demand forecasts, and then the robust voltage control problem (Sec.~\ref{sec:Robust_control}) is solved. The power setpoints from the controller are then sent to the PV inverters. Their cycles are repeated every 30 seconds till the end of the day's operation.
\subsubsection{Performance metrics}
This section defines the metrics used in the performance assessment of the estimation scheme. 
\begin{table}[!ht]
    \centering
            \caption{Performance metrics}
    \begin{tabular}{|c|c|}
    \hline
         \textbf{Metrics} & \textbf{Expression}  \\
         \hline
         RMSE$(\mathbf{\mathbf{\widehat{X}}})$ & $\frac{||\mathbf{X}^{\text{true}} - \mathbf{\widehat{X}}||_2}{||\mathbf{X}^{\text{true}}||_2}$\\
         \hline
         PICP & $\frac{1}{M}\sum_{t_k=t_1}^{t_M}b_{t_k}$ where $b_{t_k}$ counts number of times \\ 
         & the true coefficients are within the uncertainty bound.\\
   \hline
        PINAW & $\frac{1}{N (K^P_{ij,\text{max}})}\sum_{t_k=t_1}^{t_M} (2\Delta{K^P_{ij,t_k}})$\\
        \hline
        CWC & $\text{PINAW}(1+\eta \text{(PICP)} e^{-(\nu(\text{PICP} - \alpha))}$ \\
        \hline
    \end{tabular}
    \label{tab:perf_metrics}
\end{table}
The metrics are listed in Table~\ref{tab:perf_metrics}. Here, the first one is the classical
root-mean-square-error (RMSE). Here, the vectors $\mathbf{X}^{\text{true}}$ and  $\mathbf{\widehat{X}}$ contain true and estimated values of particular sensitivity coefficients, respectively for all time steps.

For the performance comparison on the estimation of the uncertainty intervals, we use metrics inspired by \cite{khosravi2013prediction}:
the first is the prediction interval coverage probability (PICP) that counts the number of instances of realization falling within the uncertainty bounds for a given confidence interval $\alpha$. 
The second is \textcolor{black}{the} prediction interval normalized average width (PINAW) to quantify the uncertainty width.
Here, $K^P_{ij,\text{max}}$ is the maximum value of the coefficient in the series.
The final metric is \textcolor{black}{the} coverage width-based criterion (CWC), which quantifies the trade-off between high PICP and small PINAW where 
\begin{small} $\eta = \begin{cases}
            0, & \text{PICP} \leq \alpha \\
            1, & \text{otherwise}
    \end{cases}$.
\end{small}
The symbol $\nu$ is \textcolor{black}{a design parameter to amplify the instances when PICP is higher than the confidence interval, based on a trade-off between the interval width penalization}. We chose $\nu = 50$. The \textcolor{black}{considered} $\alpha$ is 99\% \textcolor{black}{reflecting the confidence interval used in the estimations}. 
\subsection{Experimental Results}
The control aims to keep the voltage within 0.96 - 1.04~per unit (pu) of the base voltage. The PV inverters are controlled with a time resolution\footnote{This time resolution is chosen based on the time taken to gather the measurements, execute the estimation, update the short-term forecast of the PV production and demand, and solve the robust control problem.} of 30 seconds. The control scheme was validated for several days, but for the sake of brevity, we show for a single day; it corresponds to a weekday  (Monday, 18 July 2022). The day is characterized by clear-sky irradiance. 
The estimation and control results are described below.
\begin{figure}[!ht]
\centering
\subfloat[$K^p_{9,3}$]{\includegraphics[width=0.9\linewidth]{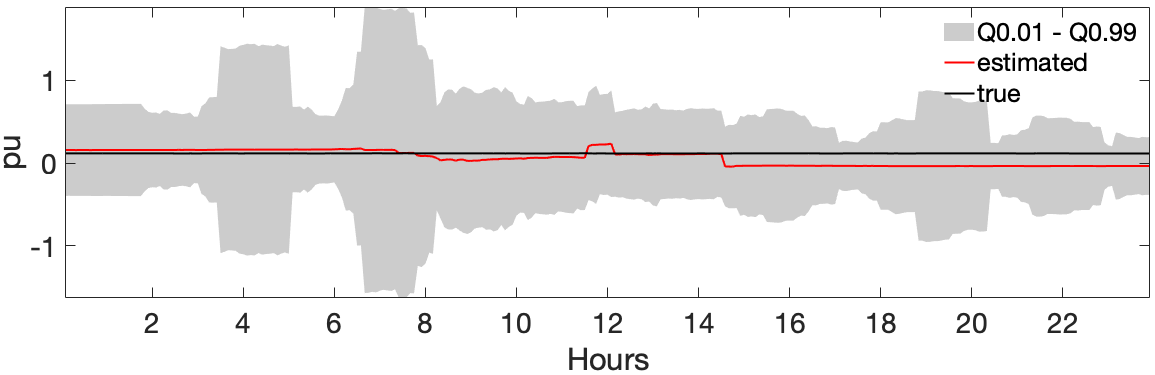}
\label{fig:est1_day2}}\\
\subfloat[$K^p_{9,9}$]{\includegraphics[width=0.9\linewidth]{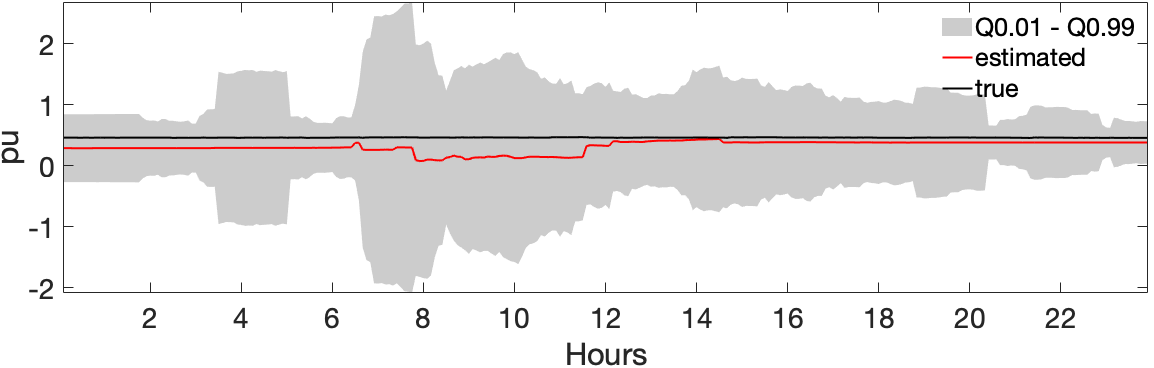}
\label{fig:est2_day2}}\\
\subfloat[$K^p_{11,3}$]{\includegraphics[width=0.9\linewidth]{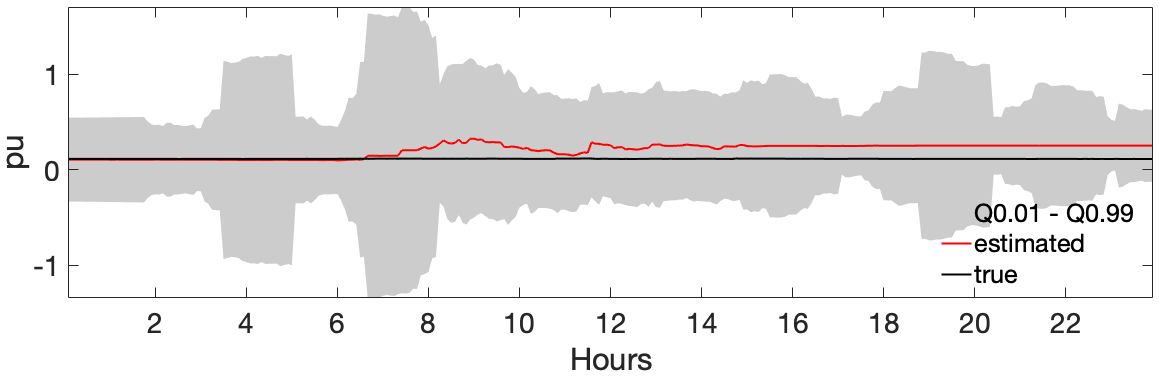}
\label{fig:est3_day2}}\\
\subfloat[$K^p_{11,11}$]{\includegraphics[width=0.9\linewidth]{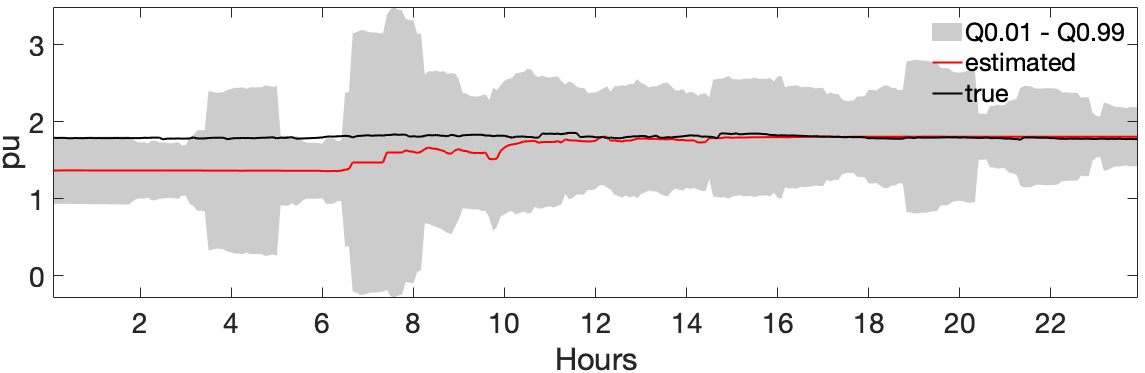}
\label{fig:est4_day2}}
\caption{Coefficients estimates and their uncertainty using RLS-SF.} \label{fig:day2_estimation}
\end{figure}
\subsubsection{Estimation Results} In Fig.~\ref{fig:day2_estimation},
we show the estimation results for nodes with controllable PV plants (nodes B09 and B11), i.e., $K^p_{9,9}$, $K^p_{9,3}$, $K^p_{11,3}$ and $K^p_{11,11}$, The estimates are shown in red, and the uncertainty on the estimates in shaded grey. They are compared against the true values (in black) obtained by model-based\footnote{Indeed, the microgrid is equipped with a real-time state estimator coupled with the method in \cite{christakou2013efficient}; this information is used for computing the true sensitivity coefficients.} computation of the sensitivity coefficients \cite{christakou2013efficient}. 
The key metrics on the RMSE and PICP-CWC-PINAW are shown in Table.~\ref{tab:metric_day2}. 
From the plots and the reported metrics, all the coefficients \textcolor{black}{except $K^p_{11,11}$} attain
nearly 100~\% coverage, although the CWC is relatively high. \textcolor{black}{The coefficient $K^p_{11,11}$ estimates are at the edge during the beginning of the day, it is due to insufficient variation in the PV injection as the PV plant is off during the night. 
This issue can be effectively tackled by increasing the confidence intervals from 99~\% to 99.9\%. In this specific case, we keep it to 99~\% as during the night there is no overvoltage problem as PVs are not generating.} 
Overall, it can be concluded that the true coefficients fall within the estimated uncertainty bounds; thus, they can be used reliably for real-time voltage control.
\begin{table}[!h]
    \centering
    \caption{Estimation performance.}
    \begin{tabular}{|c|c|c|}
    \hline 
         \bf{Coefficients} & \bf{RMSE} & \textbf{PICP-CWC-PINAW}  \\
         \hline
         $K^p_{9,3}$ & 0.88 & 1 - 11.75- 11.75\\
         \hline
        $K^p_{9,9}$ & 0.37 & 1 - 3.60 - 3.60\\
        \hline 
        $K^p_{11,3}$ & 0.95 & 1 - 11.03-11.03\\
         \hline
        $K^p_{11, 11}$ & 0.13 & 0.9 - 0.78 - 1.44\\
        \hline 
    \end{tabular}
    \label{tab:metric_day2}
\end{table}
\begin{figure}[!h]
\centering
\subfloat[Nodal voltage magnitudes with control (line plot) and without control (shaded grey).]{\includegraphics[width=0.97\columnwidth]{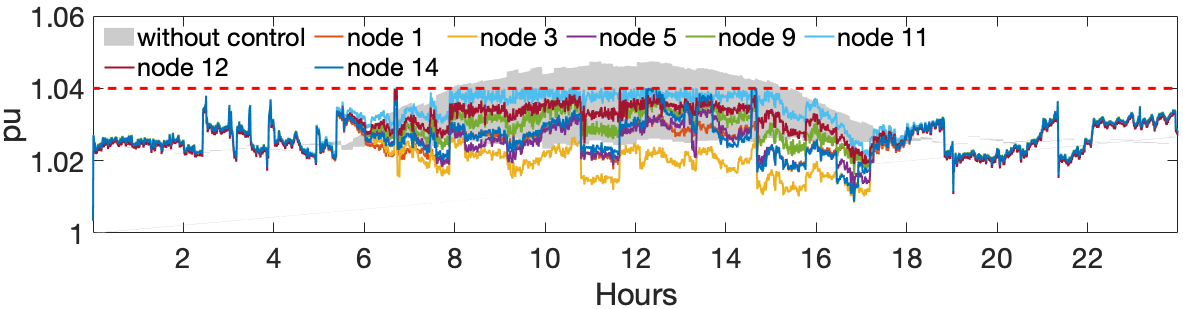}
\label{fig:Vmag_day2}}
\hfil
\subfloat[PV plant at node B11: curtailed generation (line plot) and MPP (shaded grey).]{\includegraphics[width=0.97\columnwidth]{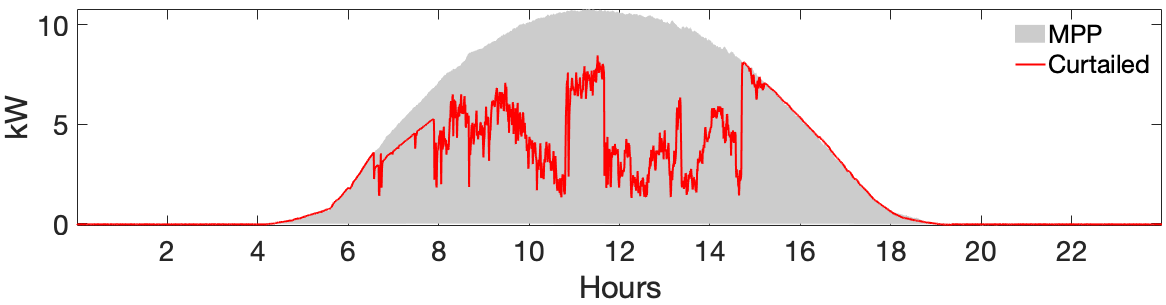}
\label{fig:PV_sm_day2}}
\hfil
\subfloat[PV plant at node B09: curtailed generation (line plot) and MPP (shaded grey).]{\includegraphics[width=0.97\columnwidth]{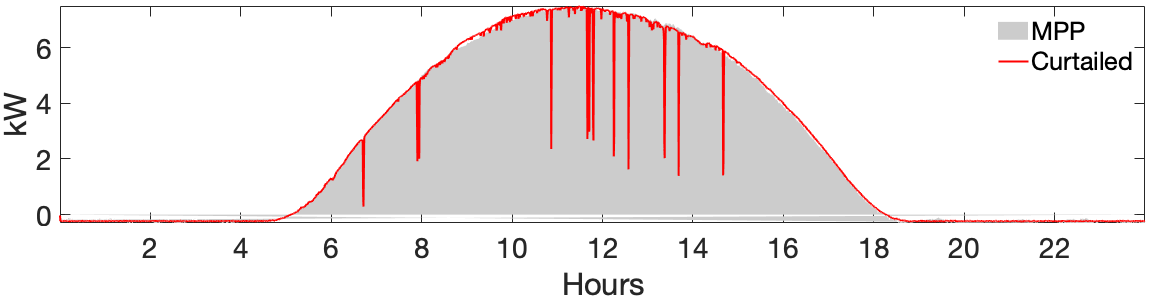}
\label{fig:PV_perun_day2}}
\hfil
\subfloat[Uncontrollable active power injections.]{\includegraphics[width=\columnwidth]{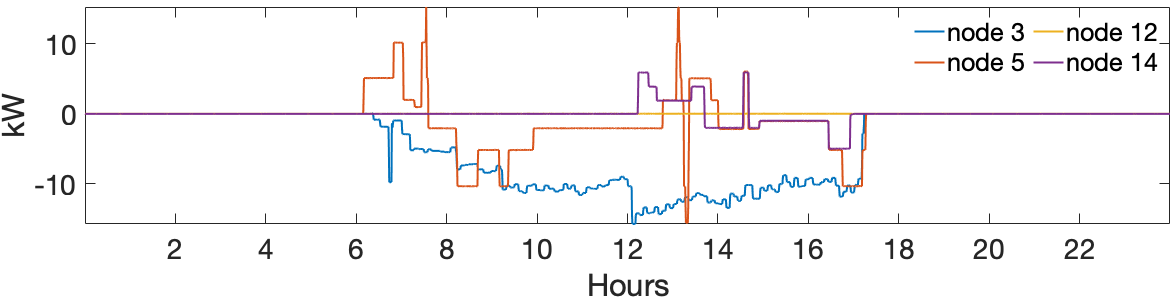}
\label{fig:Pmeas_day2}}
\hfil
\subfloat[Uncontrollable reactive power injections.]{\includegraphics[width=\columnwidth]{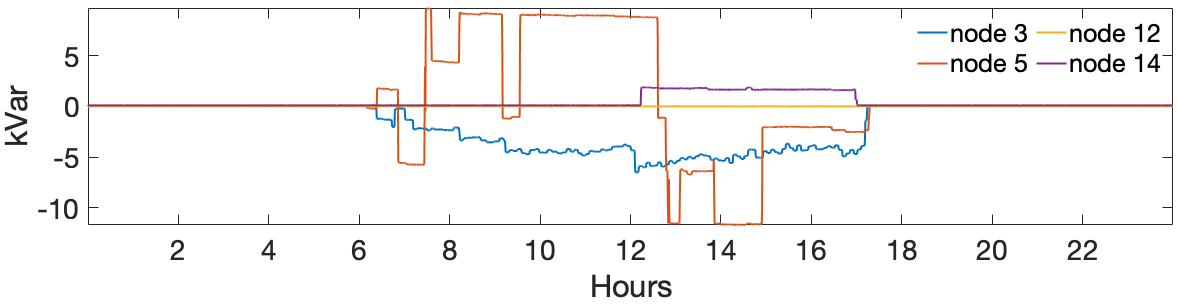}
\label{fig:Qmeas_day2}}
\caption{Experimental validation results: (a) voltage magnitude, (b) PV at node B11, (c) PV at node B09, and (d) uncontrollable active and (e) reactive powers at other nodes.} \label{fig:modelless_day2}
\end{figure}
\subsubsection{Control Results} \textcolor{black}{Fig.~\ref{fig:modelless_day2} shows the control results}. \textcolor{black}{In Fig.~\ref{fig:Vmag_day2}, it shows the nodal voltage magnitudes measurements of different nodes with and without control; the line plots show voltages with control, whereas the shaded grey area} shows the plots without any control\footnote{\textcolor{black}{As it is not possible to repeat same conditions of the experiments, this plot is obtained solving an AC load while replaying the uncontrollable power injections and imposing the maximum power potential of the PV plants.}}. \textcolor{black}{As it can be observed, all the nodal voltages are within the operational limit of 1.04~pu}, thanks to the robust control action. \textcolor{black}{Conversely, in case of no control,} the nodal voltages do not respect this limit. \textcolor{black}{Therefore, it can be concluded that robust control succeeds in voltage control. Indeed, the control action results in curtailing} PV generation from the two PV plants to keep the voltage within the imposed limit. \textcolor{black}{The curtailed PV generation and corresponding estimated maximum power potential (MPP)\footnote{\textcolor{black}{The maximum power potential of each PV plant is obtained by using a PV generation model from \cite{sossan2019solar}, fed with the measurements of the global horizontal irradiance, air temperature and the configuration of the PV plants.}} is shown in} Figs~\ref{fig:PV_sm_day2} and \ref{fig:PV_perun_day2}. Fig.~\ref{fig:Pmeas_day2} and \ref{fig:Qmeas_day2} show the uncontrollable nodes' active and reactive power injections. From the above plots, the following observations are made:
\begin{itemize}[leftmargin=*]
    \item Most curtailments occurred on the PV1 plant as it is located at the \textcolor{black}{feeder's end}, causing over-voltages across all the nodes \textcolor{black}{in case of excess} PV generation. \textcolor{black}{A fair curtailment strategy (e.g., \cite{gueissaz2017fair}) will be investigated in our future work promoting fair curtailment action.}
    \item PV plants (PV1) \textcolor{black}{experienced large curtailments between 9:00-11:00 and 12:00-15:00}; it is due to an increase in the slack's nodal voltage (imposed by the \textcolor{black}{upper-level grid.} 
    \item \textcolor{black}{The PV curtailment decreases in PV1 at 12.00 and 15.00 due to} a sudden drop in the \textcolor{black}{slack's} voltage; It is because of tap-changer's \textcolor{black}{action} in the upstream grid.
    \item \textcolor{black}{Due to the reactive power injection at node B05 between 14.00-15:00, the curtailment at PV1 plant decreases.}
\end{itemize}
\subsubsection{Computation Time}
\begin{figure}[!h]
    \centering
    \includegraphics[width = \linewidth]{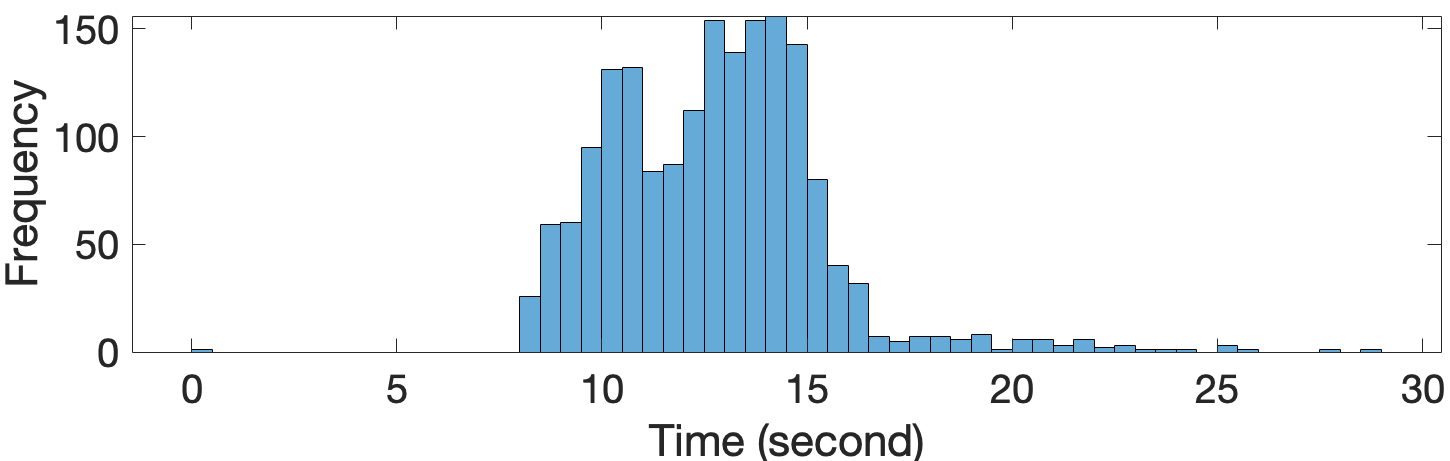}
    \caption{Probability density function (PDF) of the total computation time.}
    \label{fig:time_day2}
\end{figure}

We also report the statistics on total computation time comprised of time to download the measurements, estimate the sensitivity coefficients, obtain short-term forecasts, and solve robust control scheme. The histogram of the computation time is shown in Fig.~\ref{fig:time_day2}.
As can be observed, the whole scheme takes an average total computation time of 12.8 seconds which is below the control actuation time deadline of 30 seconds. \textcolor{black}{In some cases, the time is near the 30-second deadline\footnote{We use a fallback strategy to implement the previous setpoint in the case of exceeding the time deadline.}, it is due to the communication delay in receiving the measurements.}
\section{Conclusion}\label{sec:conclusion}
In this work, we presented an experimental validation of a model-less robust voltage control scheme on a real-life microgrid hosting two controllable PV plants. The robust control scheme relied on measurement-based estimated voltage sensitivity coefficients, and robustness is achieved by accounting for estimation uncertainties. The estimation and control problem is solved every 30 seconds, and the estimation is performed with a measurement window of 5 minutes.

The scheme was experimentally validated on a real microgrid hosted at the EPFL Distributed Electrical Systems Laboratory, a replica of the CIGRE low voltage benchmark microgrid. The control results are shown for a single day of experiments. The experimental results show that the proposed robust control scheme keeps the nodal voltage magnitudes within the imposed limits thanks to the proposed model-less robust voltage control scheme's curtailment action on the PV plants. Also, the total computation time for estimation and control was, on average, 12.8 seconds which is much below the control actuation time deadline.

Future works aim to extend this framework to realize other control objectives, such as model-less congestion management, dispatch tracking, etc.

\appendix
\subsection{Estimation of Voltage Sensitivity Coefficients}
\label{sec:sensitivity_Est}
Below we briefly describe the two-stage estimation scheme originally developed in Author's previous work in \cite{gupta2022model}.
\subsubsection{Offline LS}
\label{sec:LS}
is used to obtain initial estimates for the RLS estimation scheme, described later.
Eq.~\eqref{eq:volt_model} can be written as
\begin{align}
    |{v}_{i,t_{k}}| - |{v}_{i,t_{k-1}} | = |\underbrace{\Delta{v}_{i,t_k}}_{\gamma_{t_k}}| \approx  \underbrace{[\Delta\mathbf{p}_{t_k} ~ \Delta\mathbf{q}_{t_k}]}_{h_{t_k}}
    \underbrace{\begin{bmatrix}
        \mathbf{K}^p_{i,t_k}\\
        \mathbf{K}^q_{i,t_k}
\end{bmatrix}}_{\mathbf{X}}.\label{eq:V_linear}
\end{align}
Given the measurements window from time $t = t_1 \dots, t_M$, and assuming that coefficients do not change\footnote{\textcolor{black}{Given, the measurements are sampled at 1-sec and the network is in the steady state (no dynamic transients), this assumption holds well.}} within the window, eq. \eqref{eq:V_linear} can be written as 
\begin{align}
    \mathbf{\Gamma} \approx \mathbf{H}\mathbf{X} \label{eq:Lmodel}
\end{align}
where, $\mathbf{{\Gamma}} \in \mathbb{R}^{M\times1} = [\gamma_{t_1} \gamma_{t_2} \dots \gamma_{t_M}]^{\top}$, $ \mathbf{H} \in \mathbb{R}^{M \times 2N_b} = [h_{t_1} h_{t_2} \dots h_{t_M}]^{\top}$ and $\mathbf{X} \in \mathbb{R}^{2N_b\times 1}$ includes $\mathbf{K}^P_{i,t_k}$ and $\mathbf{K}^Q_{i,t_k}$.
The estimation problem is formulated as
\begin{align}\label{eq:RegularisedLS}
\small
    \widehat{\mathbf{X}} = \underset{\mathbf{X}}{\text{min}}||\mathbf{\Gamma} - \mathbf{H}\mathbf{X}||_2 + \lambda^{\text{reg}}\mathbf{X}^{\top}\mathbf{X}
\end{align}
where $\lambda^{\text{reg}} \geq 0$ serves as a regularization parameter to penalize coefficients assuming large values. 
It can be solved as
\begin{align}
\small
    & \widehat{\mathbf{X}}_{t_0} = (\mathbf{H}^{\top}\mathbf{H} + \lambda^{\text{reg}} \mathbf{I})^{-1}\mathbf{H}^{\top}\mathbf{\Gamma} = (\mathbf{R}_{t_0}+ \lambda^{\text{reg}}\mathbf{I})\mathbf{H}^{\top}\mathbf{\Gamma} 
    \label{eq:LSsol}
\end{align}
where $\mathbf{I}$ is the identity matrix. The covariance matrix is $\mathbf{P}^\text{cov}_{t_0} = \mathbf{R}_{t_0}^{-1} = (\mathbf{H}^{\top}\mathbf{H})^{-1}$.
\subsubsection{Online RLS} 
\label{sec:RLS}
is used to update the estimates recursively with more recent measurements during real-time operation. The scheme is initialized with LS estimates and solved recursively during the day.
A forgetting factor $0 < \mu \leq 1$ is applied to propagate covariance information from the last step as
\begin{align}
     & \mathbf{R}_{t_k} = \mu\mathbf{R}_{t_{k-1}} + h_{t_k}^{\top}h_{t_k}
\end{align}
This results in the following iterative updates. \label{eq:RLS-F}
\begin{subequations}
\begin{align}
\small
        & e_{t_k} = \gamma_{t_k} - h_{t_k}\widehat{\mathbf{X}}_{t_{k-1}}\\
        & \widehat{\mathbf{X}}_{t_k} = \widehat{\mathbf{X}}_{t_{k-1}} + \mathbf{G}_{t_k}e_{t_k}\\
        & \mathbf{G}_{t_k} = \frac{\mathbf{P}^\text{cov}_{t_{k-1}}h^{\top}_{t_k}}{\mu + h_{t_k}\mathbf{P}^\text{cov}_{t_{k-1}}h^{\top}_{t_k}}\\
        & \mathbf{P}^\text{cov}_{t_{k}} = (\mathbf{I} -  \mathbf{G}_{t_k}h_{t_k})\mathbf{P}^\text{cov}_{t_{k-1}}/\mu \label{eq:RLSFd}
\end{align}
\end{subequations}
where, $ \mathbf{G}$  {is} the estimated gain and $e$ the residual. This scheme is referred {to as} RLS-F. However, it suffers from the windup\footnote{The windup problem occurs when the system has very low excitation (i.e., the system is slowly varying); it leads to the exponential growth of the covariance matrix.} problem of the covariance matrix \cite{parkum1992recursive, vahidi2005recursive} and it may lead to very large covariances resulting in {large} estimate variances. 
A way to solve this issue is to  use different forgetting factors for different eigenvalues of the covariance matrix. These forgetting factors are computed and updated iteratively to limit the windup problem. This scheme is called selective forgetting (SF) i.e., RLS-SF; the gain and covariance matrices {are} updated as follows \cite{parkum1992recursive}.
\begin{subequations}
\begin{align}
\small
       & \mathbf{G}_{t_k} = \frac{\mathbf{P}^\text{cov}_{t_{k-1}}h^{\top}_{t_k}}{1 + h_{t_k}\mathbf{P}^\text{cov}_{t_{k-1}}h^{\top}_{t_k}}\\
        & \mathbf{P}^\text{cov}_{t_{k}} = \sum_{i=1}^{2N_b}\frac{\tau_{i,t_k}}{\mu_i}u_{i,t_k}^{\top}u_{i,t_k} \label{eq:SFe}.
\end{align}
Here, $u_{i,t_k}$ denotes the eigenvectors of $\mathbf{P}^\text{cov}_{t_k}$ in Eq. \eqref{eq:RLSFd} and $ \tau_{i,t_{k}}$ {the corresponding} eigenvalues. It is updated as $\tau_{i,t_{k}} = 
        \begin{cases}
                1, & \tau_{i,t_{k}}>\tau_{\text{max}}\\
                \tau_{\text{min}} + (1 - \tau_{\text{min}}/{\tau_{\text{max}}})\tau_{i,t_{k-1}} & \tau_{i,t_{k-1}} \leq \tau_{\text{max}}
        \end{cases} $
\end{subequations}
and bounded by $[\tau_{\text{min}}~ \tau_{\text{max}}]$.
More information on the tuning of RLS-SF is in \cite{parkum1992recursive} and \cite{fortescue1981implementation}.
\subsection{Robust Control Formulation}
\label{sec:Robust_control_append}
Without loss of generality, we consider that the distribution network is hosting controllable PV plants indexed by $j$ contained in set $\mathcal{N}_{\text{pv}} \subset \mathcal{N}_b$. The objective is to control active/reactive power injections ($p_{j,t_k}^{\text{pv}}, q_{j,t_k}^{\text{pv}}, j\in\mathcal{N}_{\text{pv}}$) such that the grid nodal voltages are always within the statutory bounds. At the same time, it minimizes the active power curtailment with respect to maximum active power potential (MPP) $\widehat{p}_{j,t_k}^{\text{pv}}$ and the corresponding reactive power given by the PV plant fixed power factor. The objective we minimize at time $t_k$ is
\begin{subequations}\label{eq:non-Robust}
\begin{align}
    \underset{{p}_{j,t_k}^{\text{pv}}, {q}_{j,t_k}^{\text{pv}}, \forall j\in\mathcal{N}_{\text{pv}}} {\text{minimize}}~ \sum_{j\in\mathcal{N}_{\text{pv}}}\Big\{(p_{j,{t_k}}^{\text{pv}}-\widehat{p}_{j,t_k}^{\text{pv}})^2  + (q_{j,t_k}^{\text{pv}})^2 \Big\} 
\end{align} 
where the first and second terms are on minimizing the active power curtailments and regulating corresponding reactive power within the PV plant's minimum power factor limit, respectively. 

The problem is solved with respect to the following constraints: 
\begin{align} 
    & 0 \leq p_{j,t_k}^{\text{pv}} \leq  \widehat{p}_{j,t_k}^{\text{pv}} \label{eq:PVp_limit_ch6} && j \in \mathcal{N}_\text{pv}\\
    & 0 \leq (p_{j,t_k}^{\text{pv}})^2 + (q_{j,t_k}^{\text{pv}})^2 \leq  ({S}^{\text{pv}}_{j,\text{max}})^2 && j \in \mathcal{N}_\text{pv},  \label{eq:capability_PV_ch6}\\
    &  q_{j,t_k}^{\text{pv}} \leq p_{j,t_k}^{\text{pv}}\zeta && j \in \mathcal{N}_\text{pv}\label{eq:pf1_ch6}\\
    &  -q_{j,t_k}^{\text{pv}}  \leq p_{j,t_k}^{\text{pv}}\zeta && j \in \mathcal{N}_\text{pv}\label{eq:pf2_ch6}\\
    &    {v}^{\text{min}} \leq |{v}_{i,t_{k}}| \leq {v}^{\text{max}}\label{eq:volt_const}.
\end{align}
Here, \eqref{eq:PVp_limit_ch6} refer to the constraint on PV generation limited by short-term MPP forecast $\widehat{p}_{j,t_k}^{\text{pv}}$, 
\eqref{eq:capability_PV_ch6} is the capability constraint of the converter rating ${S}^{\text{pv}}_{j,\text{max}}$. Eqs. \eqref{eq:pf1_ch6} and \eqref{eq:pf2_ch6} are the minimum power factor constraint (for simplicity, we assumed that all the PV plants have the same minimum power factor). Here, $\zeta = \sqrt{(1-\text{PF}^2_{\text{min}})/\text{PF}^2_{\text{min}}}$, $\text{PF}_{\text{min}}$ being the minimum power-factor allowed for the PV operation {of each PV plant}.
The final constraints (eq.~\ref{eq:volt_const}) are on the voltage magnitudes, expressed using \eqref{eq:volt_model}, are bounded by [${v}^{\text{min}}, {v}^{\text{max}}$]. 
\end{subequations}

Note that the interval constraints of \eqref{eq:intervalCoeff} used for expressing voltage magnitudes in \eqref{eq:volt_model} make {the optimization problem} in \eqref{eq:non-Robust} intractable. Therefore, the problem is reformulated using the technique proposed in \cite{ bertsimas2004price}. This approach introduces auxiliary variables $z_i, g_{ij}, y^p_j, y^q_j, j\in \mathcal{N}_\text{pv}, i \in \mathcal{N}_b$ and reformulates the constraint in \eqref{eq:intervalCoeff} and \eqref{eq:volt_const} by following set of constraints.
\begin{subequations}\label{eq:RobustVcontrol}
\begin{align}
\small
& \begin{aligned}
    |{v}_{i,t_{k-1}}| + & \Delta{\mathbf{p}_{t_k}}\widehat{\mathbf{K}}^p_{i,t_{k-1}} + 
    \Delta{\mathbf{q}_{t_k}}\widehat{\mathbf{K}}^q_{i,t_{k-1}} +  z_i\xi_i + \\ & \sum_{j\in \mathcal{N}_\text{pv}}g_{ij} \leq v^{\text{max}} ~~ \forall i \in \mathcal{N}_b\\
\end{aligned}\\
& \begin{aligned}
    |{v}_{i,t_{k-1}}| + &  \Delta{\mathbf{p}_{t_k}}\widehat{\mathbf{K}}^p_{i,t_{k-1}} + 
    \Delta{\mathbf{q}_{t_k}}\widehat{\mathbf{K}}^q_{i,t_{k-1}} -   z_i\xi_i - \\ & \sum_{j\in \mathcal{N}_\text{pv}}g_{ij} \geq v^{\text{min}} ~~ \forall i \in \mathcal{N}_b\\
\end{aligned}\\
    & -{y}^p_j \leq \Delta{{p}_{j,t_k}^{\text{pv}}} \leq {y}^p_j~~\forall j \in \mathcal{N}_{\text{pv}}\\
    & -{y}^q_j \leq \Delta{{q}_{j,t_k}^{\text{pv}}} \leq {y}^q_j~~\forall j \in \mathcal{N}_{\text{pv}}\\
    & z_i + g_{ij} \geq \Delta{K}_{ij, t_k}^p y^p_j~~i \in \mathcal{N}, j\in \mathcal{N}_{\text{pv}}\\ 
    & z_i + g_{ij} \geq \Delta{K}_{ij, t_k}^q y^p_j~~i \in \mathcal{N}, j\in \mathcal{N}_{\text{pv}}\\ 
    & {y}^p_j, {y}^q_j, {z}_i, {g}_{ij} \geq 0~~i \in \mathcal{N}, j\in \mathcal{N}_{\text{pv}}.
\end{align}
\end{subequations}
Here, the symbol $\xi_i \in [0, |\mathcal{N}_\text{pv}|]$ is a user-defined parameter for a trade-off between the robustness and conservative-ness of the solution. 

Note that the constraint reformulation of \textcolor{black}{\eqref{eq:intervalCoeff} and \eqref{eq:volt_const}} by \eqref{eq:RobustVcontrol} makes the robust problem in \eqref{eq:non-Robust} tractable and convex given its quadratic objective and linear\footnote{Eq.\ref{eq:capability_PV_ch6} is approximated by piece-wise linear equations.} constraints. Hence it can be efficiently solved using any off-the-shelf solver.

\bibliographystyle{IEEEtran}
\bibliography{bibliography.bib}
\end{document}